Thermal contact resistance between two nanoparticles

Gilberto Domingues[1], Denis Rochais[1], Sebastian Volz[2]

[1]Département Matériaux/CEA/Le Ripault, BP 16 - 37260 Monts, France

[2]Laboratoire d'Energétique Moléculaire et Macroscopique, Combustion, Ecole Centrale Paris, Grande Voie des Vignes, 92295 France

Abstract

We compute the thermal conductance between two nanoparticles in contact based on the Molecular Dynamics technique. The contact is generated by letting both particles stick together under van der Waals attractions. The thermal conductance is derived from the fluctuation-dissipation theorem and the time fluctuations of the exchanged power. We show that the conductance is proportional to the atoms involved in the thermal interaction. In the case of silica, the atomic contribution to the thermal conductance is in the range of 0.5 to 3 $nW.K^{-1}$. This result fits to theoretical predictions based on characteristic times of the temperature fluctuation. The order of magnitude of the contact conductance is 1 $\mu W.K^{-1}$ when the cross section ranges from 1 to 10nm$^2$.

The aerospace and building industries are showing increasing interest in nanoporous materials owing to their outstanding thermal insulating properties. [1-5] The thermal conductivity of these materials can be lower than that of air, which is generally considered to be a good heat insulator, with a value of 25 $mW.m^{-1}.K^{-1}$ at ambient temperature and pressure. Observation of the structure of the material on a nanometric scale explains how this thermal insulating performance is possible. The material is composed of a serie of strings of coalesced silica nanoparticles as illustrated in Figure 1. Despite the low solid volume fraction, heat transfer chiefly occurs by conduction through the coalesced nanoparticles. The two reasons for this: i) the nanoparticles have higher thermal conductivity than air, and ii) the characteristic pore size (around 100 nm) makes convection or conduction in air negligible.[6] In these nanoporous structures, coalescence zones between nanoparticles present a primordial importance in heat transfer. They govern the magnitude of the heat flux that is exchanged between nanoparticles and explain in the same time the thermal insulating capacity of these materials despite the relatively high thermal conductivity of the solid constituent. The determination of the contact conductances appears to be an essential target that will permit to understand and monitor the heat transfer through the nanoporous material.

In this article, we plan to present an original method, based on Molecular Dynamics simulations (MD), that leads to the calculation of contact conductances between nanoparticles.

After a brief description of the Molecular Dynamics technique, we establish the methodology to calculate the thermal conductances based on the application of the fluctuation-dissipation theorem.[7]

The method to put two nanoparticles in contact by MD is shown in the second part. We then present the thermal conductances versus contact surface. These conductances are finally compared to a theoretical model.

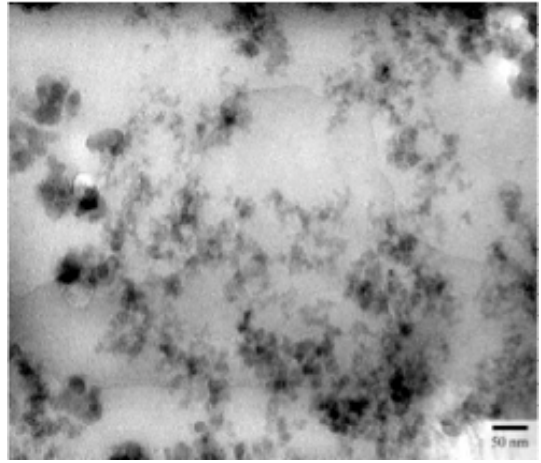

**Figure 1 :** TEM image of a nanoporous matrix of silica. Coalesced nanoparticle chains and stacks appear in dark and grey. The white zones correspond to the porosity.

## 1. Determination of the thermal interaction between nanoparticles

### 1.1 Description of the numerical approach

This section addresses the question of heat transfer on the nanoparticle scale using the MD method. This deterministic method[8] is used to describe the time trajectories of the atoms in the system by modeling them as point masses to which the fundamental law of dynamics is applied:

$$\sum_j \mathbf{f_{ij}} = M_i \ddot{\mathbf{r}}_\mathbf{i} \qquad (1)$$

where $\mathbf{f_{ij}}, M_i, \ddot{\mathbf{r}}_\mathbf{i}$ represent the interaction force between atom i and j, the mass and acceleration of atom i respectively.

Forces derive from an interaction potential which is specific to each material and which determines the accuracy of both thermal and mechanical variables. In the case of silica nanoparticles, the BKS potential was used[9] because of its accurate representation of thermal properties.[10] This potential breaks down into two terms, a Coulomb potential, which models interactions between the different charges, and a Buckingham potential, which defines the attraction and repulsion between atoms:

$$U(\mathbf{r_{ij}}) = \frac{q_i q_j e^2}{4\pi\varepsilon_0 r_{ij}} + A_{ij}\exp(-B_{ij} r_{ij}) - \frac{C_{ij}}{r_{ij}^6} \qquad (2)$$

where $q_i, q_j, r_{ij}$ correspond to the partial charges of atoms i and j, and the interatomic distance, $A_{ij}, B_{ij}, C_{ij}$ are parameters of the BKS interaction potential, $\varepsilon_0$ being the permitivity of free space.

Applying the fundamental law of dynamics to all the N atoms of the system yields a set of *3N* equations with non linear, coupled, partial derivatives, which are integrated to obtain the velocity vectors $\mathbf{v_i}$ and positions $\mathbf{r_i}$ of the atoms. Integration was achieved by using a 5$^{th}$ order Gear predictor-corrector,[11] which offers a high degree of accuracy and a good stability. The MD method can be used to express the physical variables associated with the nanoparticle solely based on the knowledge of atomic accelerations, velocities and positions.

### 1.2 Calculation of the contact conductances

Contact conductance values were calculated using a procedure inspired from studies on the characterization of heat exchanges between silica nanoparticles out of contact.[7] The microscopic expression of the power generated by a nanoparticle $NP_1$ and dissipated in a second nanoparticle $NP_2$ was used:

$$Q_{1\to 2} = \sum_{\substack{i \in NP_1 \\ j \in NP_2}} \mathbf{f_{ji}} \cdot \mathbf{v_j} \qquad (3)$$

to determine the flux ultimately exchanged between two nanoparticles (NP$_1$ and NP$_2$):

$$Q_{1\leftrightarrow 2} = \sum_{\substack{i\in NP_1 \\ j\in NP_2}} \mathbf{f_{ji}} \cdot \mathbf{v_j} - \sum_{\substack{i\in NP_1 \\ j\in NP_2}} \mathbf{f_{ij}} \cdot \mathbf{v_i} \qquad (4)$$

Since the force, related to the thermal amplitude between two nanoparticles $\Delta T$, multiplied by the term $Q_{1\leftrightarrow 2}/T$ yields a power, the fluctuation-dissipation theorem can be applied[12] to those quantities. This leads to the expression of conductance between two nanoparticles, with or without contact:

$$G_c = \frac{\int_0^\infty \langle Q_{1\leftrightarrow 2}(0) \cdot Q_{1\leftrightarrow 2}(t) \rangle dt}{k_B T^2} \qquad (5)$$

where $k_B$ is the Boltzmann constant.

### 1.3 Simulation of contacts between nanoparticles

If thermal conductances were simulated at given distances between nanoparticles,[7] we proceed to the generation of contact.

Two β-cristobalite crystals were first generated and placed at various distances from each other, without applying boundary conditions. The crystals are governed by van der Waals forces. They lose their crystalline form and move closer together. Once they come in contact, a thermostat was applied for 2000 time steps of 0.7 fs each, and the exchanged power defined by expression (4) were considered for 200.000 time steps. These data are necessary to calculate contact conductances expressed in Eq. (5).

The resulting simulations reveal a large dispersion of contacts between nanoparticles which suggests also that contact conductances values are different from one structure to another as shown in Figure 2.

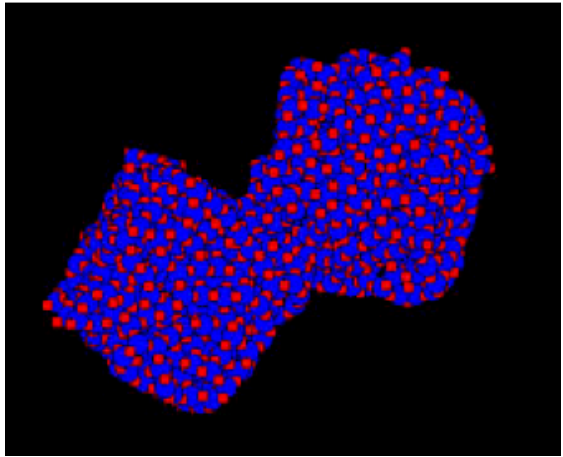

Figure 2 : Image of contact between two nanoparticles.

## 2. Results and discussion

To analyze the results according to relevant criteria, we check that the cross section is circular and defined a contact diameter and section from atomic positions. The evolution of the contact conductances with respect to calculated surfaces are presented in Figure 3. Thermal conductances are derived from Eq. (5) for nanoparticles diameters ranging from 1.5 to 5nm.

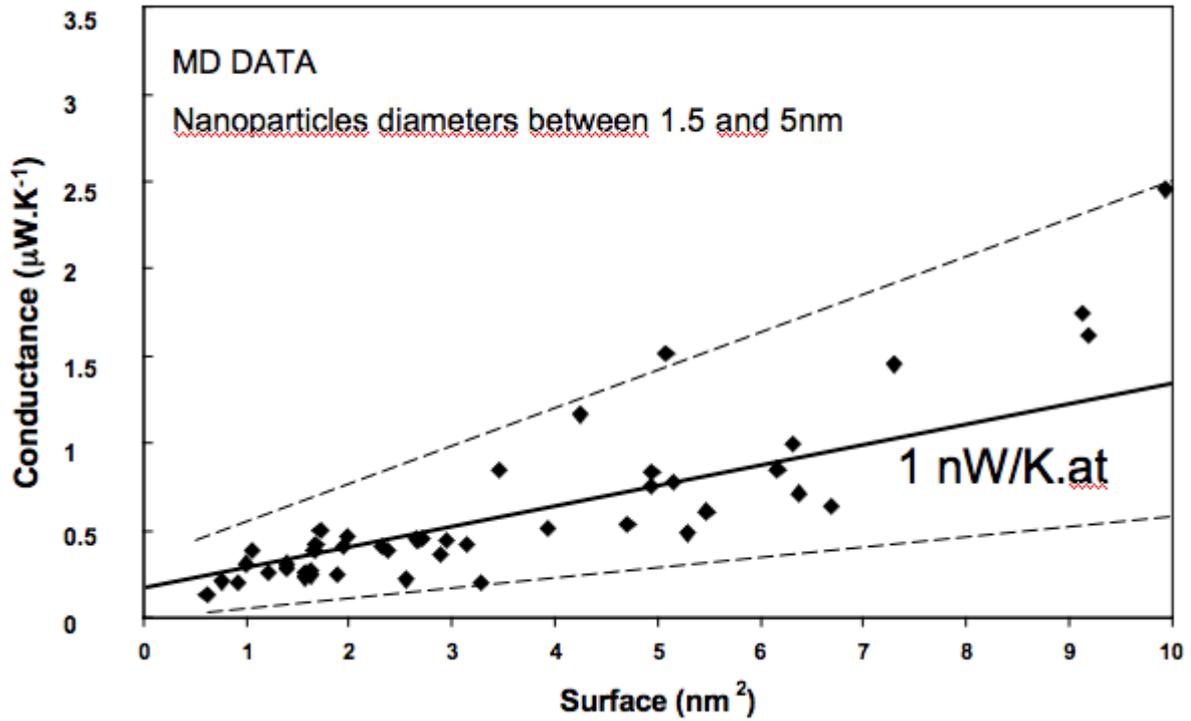

Figure 3 : Evolution of thermal conductances versus contact surface.

Figure 3 shows the dependence of thermal conductances on the contact cross section. The results for different NP diameters are reported. The increase of thermal conductance with cross section implies that atoms at contact govern the magnitude of the exchanged heat flux. Each atom can be described by a thermal conductance of value derived from the slope of the line crossing the data in Figure 3. The slope has to be divided by the cross section and multiplied by the effective surface of one atom, i.e. $10^{-20}$ m$^2$. This latter data is calculated from the interatomic distance of 0.1nm as provided by the potential. The atomic conductance is then deduced to be in the range of $g_0= 0.5\text{-}3\ nW.K^{-1}$.

The data scattering can be explained by the fact that two configurations with different NP diameters might have the same cross section. Another cause of deviation towards the linear behaviour is the difference in coordination numbers between atoms at contact. In the region of contact, the core atoms have a larger coordination number than surface atoms. Surface atoms thus have a smaller velocity and have a weaker potential interaction with the neighbouring NP. The product of the force by the velocity that is the exchanged power, is decreased. Consequently, the contribution of surface atoms to thermal conductance has to be smaller than the one of core atoms. Besides, the rate of surface atoms decreases with increasing cross section. Finally, when the section decreases, the averaged atomic thermal conductance decreases and the total conductance versus cross section deviates from the linear behaviour. This is especially true for small contact surfaces.

Comparing our simulated values with reference data is an impossible task since experimental devices are not able to measure thermal conductances at nanometric scale in irregular structures. To validate the $g_0$ value, we derive an approximated value of the atom i $G_i$ to the thermal conductance. Rewriting Eq. (4) as follows:

$$Q_{1\leftrightarrow 2} = \sum_{\substack{i \in NP_1 \\ j \in NP_2}} \mathbf{f}_{ji} \cdot (\mathbf{v}_j + \mathbf{v}_i), \tag{6}$$

and introducing this expression in the conductance leads to:

$$G_c = \frac{\int_0^\infty \sum_{\substack{i \in NP_1 \\ j \in NP_2}} \langle \mathbf{f}_{ji}.\mathbf{v}_i(0) \cdot \mathbf{f}_{ji}.\mathbf{v}_i(t) \rangle + \langle \mathbf{f}_{ji}.\mathbf{v}_j(0) \cdot \mathbf{f}_{ji}.\mathbf{v}_j(t) \rangle dt}{k_B T^2}. \tag{7}$$

The above equation was obtained by discarding the cross terms $\mathbf{v}_i$. $\mathbf{v}_j$. We now concentrate on the first term in the sum. The second term has to be on the same order of magnitude with the first because the quantities $v_i$ and $v_j$ are physically similar: atoms i and j have the same type of motion. Noting that Eq. (3) is the sum of atomic terms, the power $Q_i$ applied by NP2 on atom i can be written as:

$$Q_i \approx \mathbf{v}_i \cdot \sum_{j \in NP_2} \mathbf{f}_{ji} \approx \frac{1}{2}\mathbf{v}_i \mathbf{f}_i \approx -\frac{1}{2}\frac{\partial Ec_i}{\partial t} \qquad (8)$$

The force applied by particle 2 on atom i is assumed to be the half of the total force applied on the same atom. Indeed, the forces applied by the two NPs on contact atoms have the same modulus. The reason is that atoms of both NPs have the same separation distance with the contact atoms. Note that $Q_i$ also appears as the time variation of the potential energy of atom i in the force field of NP2. The corresponding conductance is:

$$G_i \approx \frac{1}{2}\frac{\int_0^\infty \langle \dot{Ec}_i(0)\dot{Ec}_i(t)\rangle}{k_B T^2}. \qquad (9)$$

We have multiplied $G_i$ by a factor of two to include the 2$^{nd}$ term of the sum in Eq. (7). The kinetic energy is related to the temperature as follows $Ec=(3/2)Nk_BT$. The thermal conductance $G_c$ can hence be expressed as:

$$G_c \approx k_B N^2 \frac{9}{8}\frac{\int_0^\infty \langle \dot{T}(0)\dot{T}(t)\rangle dt}{T^2}. \qquad (10)$$

where N is the number of atoms involved in the thermal interaction. Using the property of the derivative of a correlation function, it turns out that:

$$G_c \approx k_B N^2 \frac{9}{8}\frac{1}{T^2}\frac{\partial \langle T(0)T(t)\rangle}{\partial t}\bigg|_{t=0} = k_B N^2 \frac{9}{8}\frac{\langle T(0)^2\rangle}{T^2}\frac{\partial f(t)}{\partial t}\bigg|_{t=0} \qquad (11)$$

where $f(t) = \frac{\langle T(0)T(t) \rangle}{\langle T(0)^2 \rangle}$. Using the basic law of fluctuational thermodynamics $\frac{\langle T(0)^2 \rangle}{T^2} = \frac{1}{N}$, and the thermal conductance can be written in the following form:

$$G_c \approx k_B N \frac{9}{8} / \tau \qquad (12)$$

The frequency $\tau^{-1} = \frac{\partial f}{\partial t}(0)$ can be estimated by the highest frequency in the system: $\tau^{-1} \approx 40$ THz. The largest atomic conductance can therefore be estimated to $G_i = 9/8 \cdot k_B \cdot f_o \approx 0.6 \ nW.K^{-1}$ which seems reasonable when compared to the value of $g_0$ ranging between *0.5 and 3 nW.K$^{-1}$*. Another key point of Eq. (12) is that the conductance $G_c$ is proportional to the number of atoms N involved in the interaction. If the cut-off radius of the interatomic potential is small compared to the size of the nanoparticle, then the thermal conductance is essentially proportional to the number of atoms at the contact, i.e. the contact cross section. And in the limit of very large ratios of cut-off radius by nanoparticle size, the conductance is proportional to the volume of the nanoparticle. Of course, intermediate situations appear when the previous ratio is of the order of one.

## 3. Conclusion

We performed MD computations to estimate the thermal conductance between two nanoparticles in contact. The system geometry was partially random but the thermal conductance appears as increasing with contact cross section. By using a fluctuation-dissipation approach, the contribution of one atom to the contact conductance is retrieved within a satisfying accuracy compared to the value predicted by the direct calculation. The range of atomic conductance is between 0.5 and 3 nW.K$^{-1}$ whereas the theoretical prediction is 0.6 nW.K$^{-1}$. Our predictions remain approximate because the characteristic time might not be equal to the period of the optical mode. Finally, we have shown that the thermal conductance is proportional to the number of atoms involved in the thermal interaction. The conductance is therefore proportional to the contact cross section if the particle is large compared to the cut-off radius of the interatomic potential, but it is proportional to the NP volume if this cut-off is larger than the particle size.